
%
\def\parno {\par\noindent }
\magnification=1200
\hsize=14.8 true cm
\vsize=22.5 true cm
\voffset=0.1 true cm
\baselineskip=16 pt

\font\large=cmbx10 scaled\magstep1
\tolerance =5000
\vbox{\baselineskip=11 pt
\line{\hfill December 1994 } }
\vskip .9 true cm
\centerline{\large
 A Matrix Model for Random Surfaces with Dynamical Holes}

\vskip 1 true cm
\noindent
\hskip 5.2 true cm {\bf Giovanni M.Cicuta} \footnote{\dag}{
38518::CICUTA or @PARMA.INFN.IT}
\vskip .3 true cm
\centerline{
Dipartimento di Fisica, Universit\`a di Parma}
\centerline{ and INFN, Sezione di Milano}
\centerline{Viale delle Scienze, 43100 Parma}
\vskip .5 true cm
\noindent
\hskip 3.7 true cm {\bf Luca Molinari}
\footnote{\ddag}{MOLINARI@MILANO.INFN.IT}
and {\bf Emilio Montaldi}
\vskip .3 true cm
\centerline{
Dipartimento di Fisica, Universit\`a di Milano}
\centerline{ and INFN, Sezione di Milano}
\centerline{Via Celoria 16, 20133 Milano}
\vskip .5 true cm
\noindent
\hskip 5.2 true cm {\bf Sebastiano Stramaglia} \footnote{\S}{39209::STRAMAGLIA}
\vskip .3 true cm
\centerline{
Dipartimento di Fisica, Universit\`a di Bari}
\centerline{ and INFN, Sezione di Bari}
\centerline{Via Amendola 173, 70126 Bari}

\vskip .8 true cm

{\bf {Abstract.}} A matrix model to describe dynamical loops on random planar
graphs is analyzed. It has similarities with a model studied by Kazakov, few
years ago, and the O(n) model by Kostov and collaborators. The main difference
is that all loops are coherently oriented and empty. The free energy is
analytically evaluated and the two critical phases are analyzed, where the
free energy exhibits the same critical behaviour of Kazakov's model, thus
confirming the universality of the description in the continuum limit (surface
with small holes, and the tearing phase). A third phase occurs on the boundary
separating the above phase regions, and is characterized by a different
singular behaviour, presumably due to the orientation of loops.\parno

\vskip 1truecm
\vfill

{\bf 1. Introduction}\par
Field theory models with matrix-valued order parameters have been the focus
of a very large amount of investigations in the past decade. The
analysis of these models in the limit of large order of the matrices, the
large N limit, even in reduced dimension of space-time, provides important
suggestions for the non-perturbative understanding of quantum field theory
and for the formulation of string theory. \parno
Already at the beginning of the more modern developments, much attention was
given to the loop correlators
$$ \eqalign {
 W (l_1,l_2,..l_m)  =&{1\over {N^m}} \langle
 ({\rm Tr} \, e^{l_1 \Phi})\ldots ({\rm Tr} \, e^{l_m \Phi})\rangle_V =\cr
=& {1\over Z}\int  {\cal D} \Phi \,  e^{ -N {\rm Tr} \, V( \Phi)}
{1\over {N^m}} {\rm Tr} ( e^{l_1 \Phi})\ldots {\rm Tr} (e^{l_m \Phi})\cr
 }  \eqno (1.1)
$$
\noindent
and Schwinger-Dyson equations were obtained for their expectation values. The
potential $V( \Phi )$ in the measure is usually a polynomial in the hermitian
matrix variable $\Phi $. The Laplace transform of the above correlators
$$  {\tilde W}(p_1,p_2,..p_m)= {1\over {N^m}} \langle
 {\rm Tr}\left ( {1\over {p_1 -\Phi}}\right )\ldots
 {\rm Tr}\left ( {1\over {p_m -\Phi}}\right ) \rangle _V \eqno (1.2)
$$
\noindent
satisfy a chain of equations where multi-loop correlators may be evaluated from
correlators with fewer loops, and correlators of higher genus in the
topological expansion are evaluated from correlators of lower genus [1-4].
\parno
The Schwinger-Dyson equation for the one-loop correlator
$$
N \, V ^{\prime}
 ({ {\partial} \over {\partial l}}) W(l) =
 \int_0 ^l  d \, l ^{\prime}   W(l^{ \prime}, l-l^{\prime}) \eqno (1.3)
$$
\noindent
implies, in the planar limit, a quadratic equation for its Laplace transform
with a solution equivalent to the more usual saddle-point analysis.\par
The main interest of loop correlators is related to their interpretation
as insertion of loops of perimeters $l_1,\ldots l_m $ in the dynamical
triangulation
provided by the graphs dual of the planar Feynman graphs of the model. Loop
equations have a pictorial interpretation as splitting or glueing of loops
and handles.  In the continuum limit of the matrix model, obtained for
critical values of the couplings in the potential, the loop operator
may be arranged to have a finite loop length, or an infinitesimal one, thus
being referred to as a macroscopic (or respectively, microscopic)
loop operator. \par
A related analysis was done for the operator $ \psi_n = {\rm Tr}(\Phi ^n)$,
corresponding to the insertion of a n-sides polygon in the dinamical
triangulation. Duplantier and Kostov analyzed the connected two point
correlator $ \langle {\rm Tr}(\Phi ^L){\rm Tr}(\Phi ^L)\rangle _V  $,
and obtained critical coefficients for the problem of random
self-avoiding paths on a random planar graph [5]. \par
In a very interesting paper, Kazakov [6] analyzed the effects of
{\bf dynamical } loops in the simplest one matrix model, in the planar limit.
The partition function is
$$
Z = \int {\cal D} \, M \,
\exp \lbrace -N \, {\rm Tr}  [ {1 \over 2} M^2  -{g \over 4}
M^4 + L \, \log ( 1 - z^2 M^2)]  \rbrace
\eqno (1.4)
$$
(We slightly change Kazakov's notation to simplify the comparison
with the present paper). In the large N limit, the free energy
$ E= -{1 \over {N^2}} \log Z $
is the sum of planar Feynman graphs where "gluons" interact with the
quartic vertex $g$ and, in the proper continuum limit, describe
planar connected surfaces with the insertion of an arbitrary number
of holes of arbitrary lengths. The parameter $L$ may be regarded as
the hole fugacity.  By a saddle point analysis,  in the large N limit,
he showed that the model has different phases and continuum limits,
related to the value of the  parameter $g /{z^2} $ being larger or
smaller than 2/3.  The free energy is not analytic at $ L=0$  if
$ 0 < g /{z^2} < 2 / 3 $. This phase is associated to loops "eating" the
surface, the "tearing phase". \par
The same model was further on analized by Kostov [7], with the orthogonal
polynomial technique. The role of fermions in generating the dynamical loops
in matrix models of Kazakov type, eq.(1.4), was investigated by Yang [8] in
dimension one.  \par
Field theory models of surfaces with dynamical loops are
interesting for the formulation of field theories of interacting
strings. They are usually called field theories of open strings, but
have implications also for models of surfaces with handles and
no boundaries, usually referred to as field theories of closed
strings.  Indeed, after a proper identification of $ h $ couples
of boundaries, the partition function associated to a surface
with $ 2 h $ boundaries must be equal to the partition function
associated to a surface with $ h $ handles.
In the past few years large-N QCD on a generic two-dimensional
manifold has been investigated as a string model and as a topological
theory [9-12].  Some identities of the above mentioned type
were exhibited.\par
This paper is an investigation of a closely related random
matrix model which seems promising for the description of two
dimensional manifolds with oriented boundaries. The present model
is the  most straightforward analogue, in zero dimension of space-time,
of the Veneziano multi-flavour chromodynamics, with the gluon field
replaced by a Hermitian $ N \times N $ matrix $ M $ and the L-flavoured
fermions replaced by a set of L complex $ N \times N $ matrices
$ \phi_a $, $ a= 1,..L $.    \par
  The partition function of our model is
$$
Z_N (L,t,z)= \int {\cal D} M  \,   \prod_{a=1} ^L{\cal D} \phi_a   \,
\exp \{ -{\rm Tr} [ V(M) + t \sum_{a=1} ^L  \phi_a  ^{\dag}  \phi_a
- {z \over {\sqrt N}} \sum_{a=1} ^L  (M \phi_a ^{\dag} \phi_a)]  \}
\eqno (1.5a)
$$
with $z>0$ and the usual integration measures for hermitian and complex
matrices
$$
{\cal D} M = \prod_{i=1} ^N d M_{ii} \, \prod_{i<j} d \,(Re \, M_{ij}) \,
d \, (Im \, M_{ij})
\;  , \; {\cal D} \phi_a = \prod_{i,j=1} ^N  d(Re \phi_{ij})
\,  d(Im \phi_{ij})
$$
One of the two variables $t$ or $z$ is redundant, and from the beginning of
\S3 we set $t=1/2$. Our choice of potential is
$$
V(M )= {1 \over 2}M^2 + {g \over {3 \sqrt{N}}} M^3 \eqno (1.5b)
$$
\noindent
By performing the gaussian integration over the complex matrices $
\phi_a $, a rescaling in $N$ of the matrix elements $M_{ij}$, and neglecting
an irrelevant constant, the partition function (1.5a) is rewritten as
$$
Z_N (L,t,z) = \int {\cal D} M \; {\rm exp} \{ -N{\rm Tr} [ V(M) + L \,
 \log(t - zM) ] \}  \eqno (1.5c)
$$
As it is apparent from eqs.(1.5), the model, in the large $N$ limit,
describes connected planar surfaces, generated by "gluons" with cubic
interactions, with an arbitrary number of ${\bf coherently \, oriented}$
non-intersecting closed boundaries (holes), generated by the propagators of
the charged fields $ \phi_a$ (Fig. 1).

Not surprisingly this model, in the continuum limit, has a phase diagram
analogous to Kazakov's model, eq.(1.4), and almost everywhere,
the same critical coefficients.
The analysis of phase transitions in matrix models in zero dimension of
space-time performed in the past decade, shows that such phases, which only
occur at infinite N, carry several similarities with phase transitions
occurring in more realistic quantum field models, for finite N and higher
dimension of space-time.  Thus the "tearing phase" of the present toy model
seems a very promising and generic feature to be explored in more
realistic models.\par
In the Boltzmann factor in eq.(1.5a) one could add the term providing the
other orientation,
$$
 {z \over {\sqrt N}} \sum_{a=1} ^L {\rm Tr} (\phi_a ^{\dag} M \phi_a)
$$
to obtain
$$
{\rm Tr} \, \sum _1 ^L \{ t\phi_a ^{\dag} \phi_a - { z \over {\sqrt N}}
M ( \phi_a ^{\dag} \phi _a  + \phi_a \phi_a ^{\dag} ) \} =
{\rm Tr} \sum_1 ^{2L} [t \Phi_a \Phi_a - {  z\over {\sqrt N}}
M \Phi_a \Phi_a ] \eqno (1.6)
$$
\noindent
where $ \Phi_a $ is the set of $ 2L $ Hermitian $ N \times N $
matrices defined by the Hermitian and the anti-Hermitian components of
$\phi_a $. One would then obtain the partition function of the
$ O(2L) $ vector model on a random lattice [5,13-15].
In the large $N$ limit, its free energy describes connected surfaces
with any number of {\bf non-oriented}, self-avoiding loops
which {\bf are not holes}.\par
The paper is organized as follows: in section 2 we consider the $L$ expansion
of the free energy and show the relationship of its first order term with
the leading asymptotics in N of orthogonal polynomials. Next, is section 3,
we compute the free energy for the cubic interaction with charged loops and
explore, in section 4, its critical behaviour. We give a simple theorem
to show the connection of the edge behaviour of the eigenvalue density with
the critical behaviour of the parameters for its support. In the present
model the two behaviours do not coexist. The critical limit of the model
gives rise to several inequivalent continuum limits, in a way analogous
to Kazakov's model [6]. The relations between the two models and the critical
coefficients are summarized in the conclusions, section 5.
\par\vskip 1 cm

{\bf {\S2. L-expansion and orthogonal polynomial asymptotics.}}\par
In this section we discuss the loop expansion of the planar free energy
of the model (1.5c) with arbitrary potential $V(M)$, and its relation
with the asymptotics for large $N$ of the orthogonal polynomial
$P_N(\lambda ) $ with the measure $e^{-N V(\lambda )} d\lambda $.\parno
The loop expansion is the formal expansion of the free energy in powers of
L:
$$ E (L,t,z) = -\lim_{N\to\infty }{1\over {N^2}}\log Z_N =
E_0 + \sum_{k=1}^\infty  L^k E_k(t,z) \eqno {(2.1)} $$
The coefficients $E_k(t,z) $ are the generators of connected planar graphs
with $k$ holes. The further expansion of $E_k$ in powers of $z$
classifies graphs according to the total perimeter of the $k$ holes,
measured as the number of connections with the surrounding surface.\parno
We consider the first terms in (2.1), obtained by a formal expansion
in L of the partition function (1.5c). The term $E_0$ is the planar free
energy of the pure one matrix model with potential $V(M)$, solved
time ago [16]
$$ E_0 = -\lim_{N\to\infty }{1\over {N^2}}\log Z_0 \quad,\quad
Z_0=\int {\cal D}M e^{-N{\rm Tr}V(M)} \eqno {(2.2)}$$
The next term, generator of planar connected graphs with one hole, is:
$$  E_1 (t,z)=\lim_{N\to\infty} {1\over Z_0}\int {\cal D}M e^{-N{\rm Tr}V(M)}
{1\over N}{\rm Tr} \log (t-zM) = \log t -
\sum_{r=1}^\infty {{(z/t)^r}\over r} G_r^{(0)} \eqno {(2.3)} $$
where $G_r^{(0)}$ is the planar Green function with $r$ external legs of the
pure one matrix model with potential $V$. The expansion (2.3) has a simple
graphical interpretation: along the boundary of the hole
the $r$ legs of the Green function  $G^{(0)}_r$ are attached each with the
coupling  $z$; the coefficient $1/r$ follows from the rotational
symmetry. From eq. (2.3) one notes the relation between the
derivative of $E_1$ and the planar part of the one-loop correlator:
$$  {  {\partial E_1(t,1)}\over {\partial t}} ={\tilde W}^{pl}(t) $$
Given the density $\rho_0 (\lambda) $ of the pure one matrix model, $E_1$
is computed by the following integral, which will be used later
$$    E_1(t,z) = \int_{a_0}^{b_0} \,d\lambda\,
\rho_0(\lambda )\log (t-z\lambda ) \quad ,\quad t>zb_0 \eqno {(2.4)} $$
The next term $E_2$ describes surfaces with the topology of a cylinder; it
is the large N limit of the difference
$$E_2= -{1\over 2}
\langle [{\rm Tr} \log (t-zM)]^2\rangle _V + {1\over 2}
\langle  {\rm Tr} \log (t-zM)\rangle _V ^2 \eqno {(2.5)}$$
We wish now to remark that the coefficient $E_1(t,z)$ provides the
{\bf leading} asymptotic behaviour, for large N, of the orthogonal
polynomial $P_N(t)$ of the one matrix model with potential $V(M)$ in the
one arc phase. Such asymptotic behaviour has recently attracted interest,
after the works [17] where it was shown that it provides the (connected)
joint probability distributions for the eigenvalues.\parno
 According to the standard procedure, to study
the large $N$ behaviour of $Z_N$ in eq. (1.5c) one performs the change of
variables from the set $M_{ij}$ to eigenvalues $\lambda_i$ and angles,
which can be integrated. The partition function, with irrelevant constants
removed, is
$$ Z_N(L,t,z)= \int \prod_{i=1}^N d\lambda_i \prod_{i<j} (\lambda_i
-\lambda_j)^2  \exp \left \{ -N \sum_{i=1}^N \left [ V(\lambda_i) + L\log
(t -z\lambda_i) \right ]\right \} \eqno {(2.6)} $$
The monic polynomial of degree N is explicitly given by the formula [18]
$$ P_N(t)= {1\over C}  \int \prod_{i=1}^N d\lambda_i \prod_{i<j}(\lambda_i
-\lambda_j)^2 \prod_{i=1}^N (t-\lambda_i) \exp \left [
-N\sum_{i=1}^N V(\lambda _i)\right ] \eqno {(2.7)} $$
where $C$ is the proper normalization factor. The formula can be the starting
point for an asymptotics in N, with $t\in (a_0,b_0)$, as investigated by
Eynard.
To take care of the log, the relation is written as follows:
$$ P_N (t)= {1\over {2Z_0}} \left [ Z_N (-{1\over N},t+i\epsilon , 1) +
Z_N(-{1\over N},t -i\epsilon , 1) \right ]\eqno {(2.8)} $$
note that we have set $z=1$ and $L=-1/N$. For large but finite N, $\log Z_N $
is computed by means of the planar free energy (2.1)
$ E_N (-1/N,t, 1)= E_0  - {1\over N}  E_1 (t,1) +{\cal O }
({1\over {N^2}})$ where the remainder has the same weight in $N$ as the
non-planar terms of the free energy. Using the saddle point equation for the
density $\rho_0(\lambda )$
$$ V(t) - 2\int_{a_0}^{b_0} d\lambda \rho_0 (\lambda )
\log |t-\lambda | = {\rm const}
$$
we have, up to irrelevant $2\pi i$ terms and for $t$ in the support $(a_0,b_0)$
of the density:
$$  E_1 (t\pm i \epsilon,1) = {1\over 2} V(t ) \pm i\pi
\int_{a_0}^t \rho_0(s) ds  \eqno {(2.9)} $$
We then find the following leading behaviour, for large N,
consistent with the more detailed formula found by Eynard:
$$P_N (t ) \approx \exp \left \{ {N\over 2}(V(t)+c_1)\right \}
 \cos \left [ \pi N
\int_{a_0}^t \rho_0(s) ds + {\cal O}(1)\right ] \eqno {(2.10)}$$
where $c_1$ is a constant which depends on the normalization for
$P_N$ and, in particular, it vanishes for $P_N$ not monic but with
unit norm.
The omitted terms ${\cal O}(1)$ cannot be accounted for by a planar
calculation, since they would require the contribution from the graphs
on the torus and higher genera.\par
\vskip 1truecm

{\bf \S3. The free energy for connected random surfaces with holes} \par
In this section we proceed to evaluate the large  $ N $ limit of the free
energy
of the model in eqs.(1.5). Specifically, we consider in the rest of this
paper the partition function (2.6) with $t=1/2$ and, for simplicity, we only
analyze the case $g>0$.
The integral is dominated by a saddle point configuration, described by a
normalized density $ \rho ( \lambda ) $
with support $ (a,b) $, that extremizes the effective action
 $$ \eqalign { S[\rho (\lambda), C] =&
  \int_a ^b  d \lambda \;
\left [ {1 \over 2} \lambda ^2  + {g \over 3}  \lambda ^3
+ L \log (1- 2z\lambda )\right ] \rho (\lambda) +\cr
& - \int_a ^b \int_a ^b  d\lambda d \mu  \, \log|\lambda
 -\mu|  \, \rho (\lambda) \rho (\mu) + C \int_a ^b d \lambda \, \rho
(\lambda) \cr }     \eqno (3.1) $$
The parameter $C$ enforces the normalization condition. The density
is the solution of the singular integral equation
$$
   \lambda +  g\lambda ^2 -{{2zL} \over  {1 - 2z\lambda }}
- 2 P \int_a ^b  d \mu {\rho (\mu) \over {\lambda -\mu}} =0
\eqno (3.2)
$$
that is
$$
\rho (\lambda)={ 1\over {2 \pi}}   \sqrt{(b- \lambda ) (\lambda -a)}
{\big [} g \lambda  +   (g s + 1)
-z {{2s(1+gs)+ gd^2 } \over  {1   - 2z\lambda}} {\big ]} \eqno (3.3)
$$
where  $ s\equiv (a+b) / 2$ and $d\equiv (b-a) / 2 $ are determined by the
two equations
$$
 L-2= s(1+gs) {\big (} {1 \over {2z}} -s {\big )} - {{d^2} \over 2}
{\big (}1+ 3gs - {g \over {2z}} {\big )} \eqno (3.4a) $$
$$  2zL= ( s+g s^2 +  {g \over 2} d^2) \sqrt { (1-2zs)^2 -4z^2 d^2}
\eqno (3.4b)$$

 \noindent
Of course, the solution (3.3) holds if the pole $\lambda ={1\over 2z}$
is outside the support. Since we choose $z> 0$, we require
$ {1 \over {2z}} > b $.
After a long computation we obtain a simplified, yet not inspiring,
expression for the free energy
$$ \eqalign { {}&E(g,L,z)=
 -\log ({d\over 2}) + L \log \left [ {d\over 2} (h +\sqrt{h^2 -1})
\right ] - {{L^2} \over 2} \log \left [
{ {h +\sqrt {h^2 -1} }\over {\sqrt {h^2 -1}}}\right ] + \cr
{}& +L {{d^2} \over {16}} (1 +2 g s )\left [2- \left (
h - \sqrt{h^2 -1}\right ) ^2 \right ]
+ {L g \over 4} d^3 \left [ {h^3 \over 3} -{h \over 2}
-{{(h^2 -1)^{3/2}} \over 3} \right ] + \cr
{}& - {L\over 2}+ {1 \over {16 z^2}}
({g \over {3z}} +1) [1- {d^2 \over 4}(1 + 2 g s )]
+{g^2 d^6 \over {192}} + {d^4 \over{64}}[ 1+6gs+6g^2 s^2] + \cr
{}&+ {d^2 \over 4}[s(1 + g s )+ {g\over 2} d^2][{g d^2 \over{24}} +
({gs \over 6} +{1 \over 4}) ( {1 \over {2z}} +s) +
{g \over {24 z^2}} ]+\cr
{}& +{d^2 \over {48}} (1 +2gs)(2g s^3 +3 s^2 +6) +{s^2 \over 4} +
{g s^3 \over 6}\cr }\eqno (3.5)
$$
where $ h= {1 \over d} ({1 \over{2z}} -s) $ \par
As a check of the above expression, we have computed the first terms
of the Taylor expansion in L of the free energy $E(g,L,z) = E_0 (g) +
L E_1(g,z)+\ldots $. The details are given in the appendix.\par
\vskip 1cm

Before closing this section we briefly recall the simpler
model with vanishing self-interaction for the $ M$ matrix,
that is eqs. (1.5) with $ g=0$.  The analysis in the
large N limit is simpler and a few terms of the $L$ expansion of
the free energy  were evaluated long ago [19] and provides a
non trivial check
for the more involved algebra of the  present paper. \parno
When $ g=0 $ one may perform the gaussian integration of the matrix M in
eq.(1.5) to obtain, neglecting irrelevant constants,  the partition function
for a set of $L$ complex matrices $  \phi_a $ with quartic coupling
  $$
Z_N (L,z)= \int  \prod_{a=1} ^L {\cal D} \phi_a   \, {\rm exp}
 \lbrace -{\rm Tr}  [{1\over 2} \sum_{a=1} ^L  \phi_a  ^{\dag}  \phi_a
- { {z^2} \over { 2 N}} (\sum_{a=1} ^L   \phi_a ^{\dag} \phi_a  ) ^2 ]
 \rbrace \eqno (3.6)
$$
There are advantages in regarding $Z_N (L,z) $ as a model of just
one rectangular matrix $ \Phi $, of dimension $ NL \times L $
with random complex entries. In this  way the model was solved both in the
planar limit [20] and , by the orthogonal polynomial technique, in
the $ {1 \over N} $ expansion [21]. This approach clarifies an interesting
symmetry of Green functions under the exchange $L\to 1/L$, which may have a
bearing also in the present case. For square matrices, that is $L=1$, it is
well known [16] that the model model should be analyzed in the large N limit
only for $0\leq z^2 \leq z^2_{cr}=1/48 $. For generic $L$, the bound
$z^2 \leq z^2_{cr}(L)$ is the special case $g=0$ of eq. (4.9). In Fig. 2 we
plot the critical line, and exhibit the point $L=1$, $1/(2z)=2\sqrt 3 $. The
finite arc with $0<L<1$ is mapped into the infinite arc $L\ge 1$ through
the simmetry $(L,x)\Leftrightarrow (1/L,x/\sqrt L )$, where $x=1/(2z)$.\par
We remark that the saddle point eq.(3.2) is easily rewritten as a
system of two equations for the even and odd components of the
eigenvalue density $\rho (\lambda)= \rho_e (\lambda) + \rho_o (\lambda) $,
where $\rho_e(\lambda ) = \rho_e(-\lambda )$ and $\rho_o (\lambda )=-
\rho_o(-\lambda )$:
$$
g \lambda ^2 + s(1+g s) -  ({ 1 \over {2z}} -s)
  {L  \over ({ 1  \over 2z}-s)^2 -  \lambda ^2 } =
2  P \int_{-d} ^d  d y \; { u_o (y+s)  \over {\lambda -y} }
\eqno (3.7a)
$$
$$
\lambda (1 + 2gs) - { L \lambda  \over
({ 1 \over 2z} - s)^2 - \lambda ^2 }=
2  P \int_{-d} ^d  dy \; {u_e (y+s) \over {\lambda -y}}
\eqno (3.7b)
$$
As long as  $ 0 \leq z^2 \leq z_{cr} ^2 (L) $  the coupled eqs.(3.7) are
trivially equivalent to eq.(3.2). One may however consider the analytic
continuation for $ z^2 \leq 0 $ , which in the simpler model eq.(3.6)
corresponds to the usually "correct" real positive value for the quartic
coupling.  Since $ z $ would be pure imaginary, the saddle point eq.(3.2)
becomes complex and eqs.(3.7) suggest the proper path in the complex
plane, as it was evaluated in Ref. [19].\par\vskip 1cm

{\bf \S 4.  Phases and continuum limits } \par
In this section we describe the singularities of the free energy with
respect to the couplings, which lead to different phases for the model
and distinct continuum limits. \par
We recall that in one-matrix models
where the potential is a finite sum of monomials in the matrix variable,
one generally finds that such singularities are determined by the condition
that the variables specifying the support are
 not differentiable with respect
to the couplings in the potential. This in turn is equivalent to the
requirement
that the eigenvalue density should vanish at the end of the support with a
zero of order $ n + {1 \over 2} $, with the integer n larger than zero.\parno
These properties of one-matrix models are well known and were proved
using orthogonal polynomials by Itzykson and Zuber [22].
However phase transitions are more conveniently discussed by the saddle point
solution and the simple proof provided here sheds light also on the
limitations of the assertions.\par
Let us consider a polynomial potential $ V(\lambda )= \sum_k g_k \lambda^k $.
The saddle point equation for the normalized density $\rho (\lambda )$
can be solved by the Poincare'-Bertrand inversion formula [23]. In
the phase where the support is a single segment (a,b), the solution may be
written
$$
\rho (\lambda) = { 1\over { \pi}}
 {1 \over {\sqrt {(b- \lambda ) (\lambda -a)}} } {\cal P}
( \lambda , b, a, g_i)
$$
$$
{\cal P}( \lambda , b, a, g_i)=
1+ { 1\over {2 \pi}}
 P \int_a ^b  d \mu  \sqrt {(b- \mu ) (\mu -a)} \,
{V' (\mu) \over { \mu - \lambda }}
\eqno (4.1)
$$

\noindent
$ {\cal P}( \lambda , b, a, g_i) $ is a polynomial in the variable
$ \lambda $ whose coefficients are entire functions of  $b $, $ a$ and $ g_i $.
The end points  $ a (g_i), b(g_i) $ of the support  are  determined by the
conditions
$$
{\cal P} ( \lambda = a, b, a , g_i)= {\cal P} ( \lambda = b, b, a , g_i)= 0
\eqno (4.2)
$$
Inserting the identity
$$ { {\sqrt {(b-\mu )(\mu- a)} }\over {\mu -\lambda }}=
{ {(b-\lambda )(\lambda -a)}\over {(\mu -\lambda )\sqrt {(b-\mu )(\mu- a)} }} +
{ {a+b-\lambda -\mu }\over {\sqrt {(b-\mu )(\mu- a)} }}$$
in eq. (4.1) and imposing the conditions (4.2),
one obtains the following equations for the extrema
$$ {1\over {2\pi }} \int_a^b d\mu  {{V'(\mu)}\over {\sqrt {(b-\mu )(\mu -a)}}}
=0 \quad ,\quad
{1\over {2\pi }}\int_a^b d\mu  {{\mu V'(\mu)}\over {\sqrt {(b-\mu )(\mu -a)}}}
=1 \eqno {(4.3)} $$
and the factorization
$$
{\cal P}( \lambda , b, a, g_i)= (b- \lambda ) (\lambda -a){\cal Q}
( \lambda ) \eqno (4.4)
$$
$$ {\cal Q} ( \lambda ) =
{1\over {2\pi }} \int_a^b d\mu  {{V'(\mu)-V'(\lambda )}\over {\mu-\lambda }}
{1\over {\sqrt {(b-\mu )(\mu -a)}}} $$
The above formula implies that, generically, the density $\rho (\lambda) $
vanishes with a zero of order one half at the extrema of its support. \par
Since the free energy $ E(g_i) $ may be evaluated as a polynomial functional
of $ \rho (\lambda) $ and because of the polynomial nature of
$ {\cal Q}( \lambda, b, a, g_i) $, the singularities of $E(g_i)$
may only occur as singularities of the functions $ a(g_i) $
and $ b(g_i) $.
By differentiating eqs.(4.2) with respect to any free parameter $g_i$, and
carrying the computations of the required partial derivatives in the parameters
$a$, $b$ on formula (4.1) and in the variable $\lambda $ on formula (4.4),
one eventually finds
$$
{ \partial a \over {\partial g_i}}= - 2
{ {\left (  { \partial {\cal P} \over {\partial g_i}} \right )_{\lambda =a}}
\over { (b-a) {\cal Q} (a)} } \quad , \quad
{ \partial b \over {\partial g_i}}= 2
{ {\left (  { \partial {\cal P} \over {\partial g_i}} \right )_{\lambda =b} }
\over { (b-a) {\cal Q} (b)} }
\eqno (4.5)
$$
Therefore, the singularities of the left sides of eqs.(4.5) may only occur at
the zeros of the denominators of the right sides. These in turn imply a
non-generic order for the vanishing of the density at the edge of
its support. \par
 We recall important examples where the above argument is evaded.
In the attempt to describe random surfaces with extrinsic curvature,
matrix models were proposed where the potential $ V(M)$ of the
Hermitian matrix $M$ is sum of monomials and the same invariant trace may
occur with different powers [24]. The simplest example is

$$
V(M)=  a \,Tr(M^2) +b \, Tr(M^4) + c \, [Tr(M^2)]^2 \eqno {(4.6)}
$$
The eigenvalue density is easily found in the large-N limit by
the saddle point analysis. However, more parameters determined by
more equations occur, and the general features above described in
eqs. (4.1-4) must be generalized. Indeed it was shown that these
models yield susceptibilities
with unusual critical coefficients [24,25]. \par
A second class of models which escape the above theorem occur
if the one-matrix potential is not a polynomial. This is the
case of Kazakov model and of the present paper. Then the function
$ {\cal Q}( \lambda,b, a, g_i)$ in eq.(3.3) is not a polynomial
and the singularities in eq.(3.4) may rise from singularities of
the numerator in the right side of the equation.
\par
We proceed to analize the critical behaviour of our model.
\vskip 1cm

To investigate the singular behaviour of the end-points $a$ and $b$ of
the support, or equivalently of the functions $s$ and $d$ given by eqs. (3.4),
it is convenient to introduce the new variables
$$ \sigma \equiv gs \quad ,\quad \delta\equiv gd \quad,\quad \tau\equiv
 {g\over {2z}} \eqno {(4.7)}$$
{}From eq. (3.4a) we isolate $\delta $
$$ {{\delta^2}\over 2}={{\sigma (1+\sigma )(\tau - \sigma )-g^2(L-2)}\over
{1+3\sigma -\tau }} \eqno {(4.8a)} $$
which allows to rewrite (3.4b) as a single equation for $\sigma (g^2,L,\tau)$:
$$ g^2L={{\sigma (1+\sigma)(1+2\sigma )-g^2(L-2)}\over {1+3\sigma -\tau }}
\sqrt { (\tau-\sigma )^2 - 2{{\sigma (1+\sigma )(\tau - \sigma )-g^2(L-2)}\over
{1+3\sigma -\tau }} } \eqno {(4.8b)} $$
A special situation occurs for $1+3\sigma -\tau =0$, and will be discussed
later. The above formulae are also the starting point for the expansion in $L$
of the free energy, as discussed in the Appendix.\parno
The singular behaviour of the function $\sigma (g^2,L,\tau )$ can be
characterized by the condition $\partial (g^2)/ \partial \sigma =0$ which,
by eq. (4.8b), provides a constraint on the parameters. Together, the
equations describe in the space $(g^2,L,\tau )$, surfaces of criticality
$g^2 = g^2_{cr}(L,\tau )$. Such surfaces will be now investigated in the form
of $L$ expansions. We could equally well consider $L$ and $g^2$ as spectators,
and require the condition $\partial \tau /\partial \sigma =0$, which would
provide an identical equation for the critical behaviour. The second
equation is:
$$ 3g^2 L = (6\sigma^2+6\sigma +1)\sqrt {(\tau-\sigma )^2 - \delta^2 }+
{{g^2L(1+3\sigma-\tau)}\over {(\tau-\sigma )^2 - \delta^2 }} \cdot $$
$$\cdot \left [
\sigma -\tau - {{(1+2\sigma )(\tau - \sigma )-\sigma (1+\sigma )}
\over {1+3\sigma -\tau }} + 3 {{\sigma (1+\sigma )(\tau - \sigma )
+ g^2(L-2)}\over {(1+3\sigma -\tau)^2}} \right ] \eqno {(4.9)}$$
In a way fully analogous to Kazakov model [6] we find
three phases for the continuum limit for small values of $ L$.\par

1. {\bf The small holes phase}. If $ \tau > {1 \over 2} (\sqrt{3} -1 )
$ the set of eqs.(4.8) and (4.9) allow a Taylor expansion
$$\sigma (\tau) = \sigma_0 + L \, \sigma_1 (\tau) + L^2 \sigma_2 (\tau)+...$$
$$\delta^2 (\tau) = \delta_0 ^2 + L \, \delta_1 ^2 (\tau) + L^2
    \delta_2 ^2 (\tau) +...$$
$$g_{cr} ^2 (\tau) = g_0 ^2 + L \, g_1 ^2 (\tau) + L^2 g_2 ^2 (\tau) +...
\eqno (4.10)$$
\noindent
where
$$\sigma_0 ={ {-3 + \sqrt{3}} \over 6} \quad ; \quad \delta_0 ^2 ={1 \over 3}
 \quad ; \quad
g_0 ^2 = {1 \over {12 \sqrt{3}}}$$
$$g_1 ^2 = {1 \over {24 \sqrt{3}}} \left [1 - \sqrt { { {2 \tau +1 - \sqrt {3}}
\over { 2 \tau +1 + {1 \over \sqrt{3}} } } } \right ]$$
$$\sigma_1 = {1 \over {36}} [ \tau + {  {\sqrt{3} +1} \over 2}]
[\tau - { {\sqrt{3} -1} \over 2}]^{-{1 \over 2}}
[\tau + { {\sqrt{3} +3} \over 6}]^{-{3 \over 2}}
\eqno (4.11)$$
The expansions (4.10) are supposed to hold for $ L \leq L_{cr} $.
Since the criticality is very similar to the pure gravity matrix
model, we expect that the free energy behaves as
$$ E (L) \sim_{ L\to L_{cr}} ( L- L_{cr} )^{ {3 \over 2}}\eqno (4.12)$$

2. {\bf The tearing phase}. If $ 0< \tau < {1 \over 2}(\sqrt{3} -1)$
the eqs.(4.8),(4.9) imply a non analytic contribution for
$L$ close to zero. We find
$$ \sigma(\tau) = \sigma_0 (\tau) + L^{2/3} \sigma_1 (\tau) +... $$
$$ \delta^2 (\tau) = \delta_0 ^2 (\tau) + L^{2/3} \delta_1 ^2 (\tau) +... $$
$$ g_{cr}^2 (\tau )=g_0 ^2 (\tau)+L^{2/ 3} g_1 ^2 (\tau)+...\eqno (4.13)$$
\noindent with
$$ \sigma_0 (\tau) = {1 \over 3}[\tau -1 + \sqrt{ 1 - 2 \tau - 2 \tau ^2}]$$
$$ \delta_0 ^2 (\tau) =-2 \sigma_0 (\tau) [ 1 + \sigma_0 (\tau)] $$
$$ g_0 ^2 (\tau) = -{1 \over 2} \sigma_0 ( \tau) [ 1 + \sigma_0
(\tau)][ 1 +2 \sigma_0 (\tau)]$$
\noindent
The reality of the roots implies the above mentioned
allowed range  for $ \tau $. \par
By inserting the expansions, eq. (4.13) in the expression
of the free energy, eq.(3.5), one finds a non analytic contribution
for $ E(L) $ at $ L=0$, of the form $ E(L) \sim L^{2 \over 3} $.
\par
3. {\bf The line separating the two phases. } A third phase is found on the
line $ \tau= 1 + 3 \sigma $. The set of eqs. (4.8) allow (4.9)
very simple analysis for any value of $L$. We find
$$
54 g^4 [L^2 -8 L +8] +27 g^{ 8 \over 3} L^{4 \over 3} -1 =0
$$
$$
\sigma = {1 \over 6} [ -3 + \sqrt{ 3 + 18 (g^2 L )^{2 \over 3}}]
$$
$$
\delta ^2 = 3 - ( g^2 L)^{1 \over 3} + 18 ( g^2 L)^{2 \over 3}
\eqno (4.14)
$$
and the following expression for the density, with a non-generic edge
behaviour:
$$ \rho (\lambda ) = {{g^2}\over {2\pi }} { {(b-\lambda )^{3/2}(\lambda -a)
^{3/2} }\over {1+3\sigma -g\lambda }} \eqno {(4.15)} $$
By expanding about $L=0$, we have
$$
\sigma = \sigma_0 + L^{2 \over 3} \sigma_1 +...
$$
$$
\delta^2 = \delta_0 ^2 + L^{1 \over 3}  \delta_1 ^2 +...
$$
$$
g^2 = g_0 ^2 + L g_1 ^2 + L^{4 \over 3} g_2 ^2 +...
$$
where $\sigma_0$, $\delta_0^2$ and $g_0$ are given in (4.11). We then obtain
the dominant singular behaviour of the free energy at $L=0$:
$$ E_{sing} (L) \approx {7\over 4}\left ( {1\over {12}} -\sqrt 3 \right )
L^{1/3}  \eqno {(4.16)} $$

\vskip 1 cm
{\bf \S5.Conclusions}\par
In this paper we analyzed a model of random surfaces with boundaries, defined
in eqs. (1.5), evaluated the free energy in the spherical limit, eq. (3.5),
and the continuum limits associated to the phases of the model.\parno
It may be worth noticing that matrix models of surfaces with boundaries
discussed in the past few years have a geometrical picture as the dual graphs
of Feynman graphs generated by a potential $V(\Phi )$ defining the random
"triangulation", and the insertion of higher order vertices ${\rm Tr}\Phi^n $
defining loops with $n$ sides. This is also the case of Kazakov model eq.(1.4),
where the higher order vertices originate from the expansion of the
logarithmic term. The similarity of Kazakov's model with the present one leads
to similar phase diagrams and the same critical exponents (with the
exception of one boundary line).\parno
Fig. 3 shows the phase diagram of Kazakov's model. Critical behaviour occurs
only below the line $L=4/z^2$, with three inequivalent continuum limits. The
phase describing surfaces with small holes is for $g/z^2 >2/3$, whereas
the "tearing phase" is for $g/z^2 <2/3$. By writing the dominant singular
behaviour of the free energy in the form $E(L)\approx (L-L_{cr})^\gamma $,
Kazakov found $\gamma =3/2$, $L_{cr}>0$ for the small holes phase;
$\gamma = 2/3$, $L_{cr}=0$ for the tearing phase, and $\gamma =4/5$, $L_{cr}
=0$ on the boundary $g/z^2 =2/3$.\parno
Fig. 4 is the phase diagram of the model analyzed in this paper. The plotted
line, with equation
$$ \tau =1+3\sigma =-{1\over 2} +{1\over 2}\sqrt { 3+18 (g^2L)^{2/3} }
\eqno {(5.1)}$$
is the boundary between a "small hole phase" (right) and a "tearing phase"
(left). For the dominant singular behaviour of the free energy we obtain:
$\gamma =3/2$, $L_{cr}>0$ for the small holes phase;
$\gamma = 2/3$, $L_{cr}=0$ for the tearing phase, and $\gamma =1/3$, $L_{cr}
=0$ on the boundary. The more complicated algebra of our model did not allow
us to produce the expected upper boundary of the critical regions, analogous
to the dotted line in Fig. 3.\parno
The main difference to be remarked is the critical exponent on the boundary
in the parameter space which separates the "small holes" phase from the
"tearing" phase. In the geometrical picture described above, this lack of
universality should be related to the occurrence in the present model of loops
with even and odd number of sides.\parno
It seems however better to consider the picture associated to the planar
Feynman graphs of eqs. (1.5a,b). The random surface is then defined by
the regular trivalent graphs and the loops by the closed paths which are
sequences of propagators of the charged field. In the spherical limit the
loops are empty and the boundaries are all coherently oriented. This picture
is closer to the O(n) vector model on random surfaces [13-15], where
non-intersecting loops are drawn on trivalent planar graphs. Because those
boundaries are not oriented, they may include parts of the graph and/or
other loops. Presumably this originates the different class of universality
of the O(n) vector model.\par
\vskip 0.5cm
{\bf {Acknowledgements}} \par
One of us, G.M.C., likes to thank prof. R. Blankenbecler for the kind
ospitality at SLAC, where part of this work was done and Dr. A. Zee for
a useful conversation about ref. 17.
\vskip 1cm

{\bf \S   Appendix} \par
We here assume that the free energy, eq.(3.5), and the parameters
$\sigma= gs$, $\delta = gd $ computed by eqs.(4.8), allow a formal Taylor
expansion around $L=0$, and we quote the first term $E_1$. Setting
$$ \sigma (g,L,\tau) = \sigma _0 (g) + L \sigma _1 (g,\tau)
 + L^2  \sigma _2 (g,\tau ) + \ldots \eqno (A.1a)$$
$$ \delta (g,L,\tau) = \delta _0 (g) + L \delta _1 (g,\tau ) +
 L^2  \delta _2 (g,\tau ) + \dots  \eqno (A.1b) $$
we find, at the lowest order, the parameters of the pure cubic model:
$\sigma _0 (g) $ solves the cubic equation
$$ \sigma _0 ( 1+\sigma _0 ) ( 1 + 2 \sigma _0 ) + 2 g^2 =0 \eqno (A.2a) $$
$$ \delta _0 ^2 = -{1 \over 2}  \sigma _0 ( 1+\sigma _0 )  \eqno (A.2b) $$
At the next order we give $\sigma_1 $ only:
$$ \sigma _1 (g, \tau) = {{g^2}\over {1 + 6 \sigma _0 + 6 \sigma _0 ^2}}
{\Big [} 1+ {{( 1  -  \tau + 3 \sigma_0) } \over
\sqrt { (\tau -\sigma _0)^2 + 2  \sigma _0 ( 1+\sigma _0 ) }} {\Big ]}
\eqno (A.3) $$
The above coefficients lead to the evaluation of the first two coefficients
of the L expansion of the free energy
$$ E_0 (g) = {1 \over 2} \log [1 + 2 \sigma_0] - {1 \over 3}
  \sigma_0 {  { 2+6 \sigma_0 + 3 \sigma_0 ^2} \over
  { (1 + 2 \sigma_0)^2 (1 + \sigma_0)} } \eqno (A.4) $$
which obviously  reproduces the well known free energy of the cubic model [16],
and
$$ E_1 (g, \tau) = -2\log (2g) + \log \left [ (\tau-\sigma_0 )
+\sqrt {(\tau -\sigma_0 )^2 -\delta_0 ^2}\right ] -1
 - {1\over {6g^2}}\sigma_0^2 (3+2\sigma_0) +$$
$$+ {1\over {g^2}}({1\over 2}\tau ^2+ {1\over 3}\tau^3 ) + \left [
{1\over 3} {{\sigma_0 (2+3\sigma_0)+\tau (6\sigma_0^2+11\sigma_0+6)}
\over {(1+2\sigma_0)(1+\sigma_0)}} + $$
$$  +{1\over {6g^2}} (\tau^2\sigma_0 (1-2\sigma_0) + \tau^3 (2\sigma_0-3)
- 2\tau^4 ) \right ]
{1\over {\sqrt {(\tau -\sigma_0 )^2 -\delta_0 ^2} }}
\eqno (A.5)
$$

\vskip 1 cm
\centerline {\bf References}
\item {[1]}{S.Wadja, Phys. Rev. {\bf D24} (1981) 970.}
\item {[2]}{A.Migdal, Phys. Rep. {\bf 102} (1983) 199.}
\item {[3]}{F.David, Mod. Phys. Lett. {\bf A5} (1990) 1019.}
\item {[4]}{J.Ambjorn et al., Mod. Phys. Lett. {\bf 5A} (1990) 1753 and
            Phys. Lett. {\bf B251} (1990) 517.}
\item {[5]}{B.Duplantier and I.Kostov, Nucl.Phys. {\bf B340} (1990) 491.}
\item {[6]}{V.A.Kazakov, Phys. Lett. {\bf B237} (1990) 212.}
\item {[7]}{I.K.Kostov, Phys. Lett. {\bf B238} (1990) 181.}
\item {[8]}{Z.Yang, Phys. Lett. {\bf B257} (1991) 40.}
\item {[9]}{B.Rusakov, Mod. Phys. Lett. {\bf A5} (1990) 693.}
\item {[10]}{D.Gross, Nucl. Phys. {\bf B400} (1993) 161.}
\item {[11]}{D.Gross and W.Taylor IV, Nucl. Phys. {\bf B400} (1993), 181 and
              {\bf B403} (1993) 395.}
\item {[12]}{M.Caselle, A. D'Adda, L.Magnea and S.Panzeri, Nucl. Phys.
            {\bf B416} (1994) 751.}
\item {[13]}{I.K.Kostov, Mod. Phys. Lett. {\bf A4} (1989) 217.}
\item {[14]}{M.Gaudin and I.Kostov, Phys. Lett. {\bf B220} (1989) 200.}
\item {[15]}{I.Kostov and M.Staudacher, Nucl. Phys. {\bf B384} (1992) 459.}
\item {[16]}{E.Brezin, C.Itzykson, G.Parisi and J.B.Zuber,  Comm. Math. Phys.
             {\bf 59} (1978) 35;
             D.Bessis, C.Itzykson and J.B.Zuber, Adv. Appl. Math. {\bf 1}
             (1980) 109. }
\item {[17]}{E.Brezin and A. Zee, Nucl. Phys. {\bf B402} (1993) 613 and
             {\bf B424} (1994) 435; B.Eynard, Saclay preprint SphT/93-999.}
\item {[18]}{G.Szego "Orthogonal Polynomials", American Math. Soc. Coll.
             Pub. vol. 23, (1939).}
\item {[19]}{A.Barbieri, G.M.Cicuta and E.Montaldi, Nuovo Cim. {\bf 84A}
            (1984) 173. }
\item {[20]}{G.M.Cicuta, L.Molinari, E.Montaldi and F.Riva, J. Math. Phys.
             {\bf 28}  (1987) 1716.}
\item {[21]}{A.Anderson, R.C.Myers and V.Periwal, Phys. Lett. {\bf 254B}
            (1991) 89 and Nucl. Phys. {\bf B360} (1991) 463;
             R.C.Myers and V.Periwal,  Nucl. Phys. {\bf B390} (1993) 716.}
\item {[22]}{C.Itzykson and J.B.Zuber, J. Math. Phys. {\bf 21} (1980) 411.}
\item {[23]}{N.I.Muskhelishvili, "Singular Integral Equations", P.Noordhoff,
             (1953).}
\item {[24]}{S.R.Das, A.Dhar, A.N.Sengupta and S.R.Wadia, Mod. Phys. Lett.
            {\bf A5} (1990) 1041.}
\item {[25]}{G.M.Cicuta and E.Montaldi, Mod. Phys. Lett. {\bf A5} (1990)
             1927.}
\par\vfill\eject

\centerline {Figure Captions}
\vskip 1truecm \par

Fig. 1 \parno
A portion of a planar graph generated by the model (1.5a). The loops formed
by black arrows are generated by the propagators  of the charged fields.\par

Fig. 2\parno
The critical line $z^2=z^2_{cr}(L)$ of the model (3.6).\par

Fig. 3\parno
Phase diagram of Kazakov's model. The continuous line has equation $g/z^2=
2/3$\par

Fig. 4\parno
Phase diagram of the present model. On the horizontal axis are the values
of $\tau = g/(2z)$. The curve originates at $\tau = (\sqrt 3 -1)/2$,
corresponding to the critical values of the parameters of the one-matrix model
with cubic interaction.
\vfill\eject
\end